\documentclass[aps,prd,preprint,amsmath,superscriptaddress]{revtex4}

\usepackage{graphicx}

\begin{document}

\title{ Constraining supersymmetry from the satellite experiments }

\author{Xiao-Jun Bi}
\email{bixj@mail.ihep.ac.cn}
\affiliation{Key laboratory of particle astrophysics, IHEP, 
Chinese Academy of Sciences, Beijing 100049, P. R. China}

\begin{abstract}
In this paper we study the detectability of $\gamma$-rays
from dark matter annihilation in the subhalos of the Milky Way 
by the satellite-based experiments, EGRET and GLAST.
We work in the frame of supersymmetric extension of the standard
model and assume the lightest neutralino being the dark matter particles.
Based on the N-body simulation of the
evolution of dark matter subhalos we first calculate the
average intensity distribution of this new class of $\gamma$-ray
sources by neutralino annihilation.
It is possible to detect these $\gamma$-ray sources 
by EGRET and GLAST.
Conversely, if these sources are not detected the 
nature of the dark matter particls will be constrained by these experiments,
depending, however, on the uncertainties of the subhalo profile.
\end{abstract}

\maketitle

\section{ introduction }

In the third EGRET catalog \cite{3cat} unidentified $\gamma$-ray
sources accounted for more than a half of the discrete sources detected
by EGRET.
Despite a great effort, most of them can not be associated with the
known sources detected at low energies up to now. 
Most of the efforts in identifying
these sources focused on the Galactic counterparts, such as young pulsars
\cite{Hobbs}, microquasars \cite{paredes}, supernova remnants \cite{torres}.
At the same time, multiwavelength searches continue to look for counterparts
in the $\gamma$-ray sources, such as blazars, supernova remnants and pulsars
\cite{fegan}.
There are recent theoretical works
trying to explain the unidentified sources as annihilating dark matter
clumps within the Milky Way (MW) \cite{flix}. These efforts try to
solve the nature of dark matter.

The existence of cosmological dark matter has been firmly established by
a multitude of observations,
 such as the observations of the rotation curves in
spiral galaxies and velocity dispersion in elliptical galaxies, the
X-ray emission and peculiar velocities of galaxies in the clusters of
galaxies, the weak lensing effects,
all indicating much steeper gravitational potentials than those
inferred from the luminous matter. 
However, the nature of the non-baryonic dark matter is still unknown
and remains one of the most outstanding
puzzles in particle physics and cosmology.

Among a large amount of theoretical candidates, the most attractive
scenario involves the weakly interacting massive particles (WIMPs).
An appealing idea is that the WIMPs form the thermal relics of the
early universe and naturally give rise to the relic abundance in the
range of the observed values. The WIMPs are well motived
theoretically by the physics beyond the standard model to solve the
hierarchical problem between the weak scale and the Planck scale.
In particular, the minimal supersymmetric extension of the standard model
(MSSM) provides an excellent WIMP candidate as the lightest supersymmetric 
particle, usually the lightest neutralino, which are stable due to R-parity 
conservation \cite{jungman}. The cosmological constraints on the 
supersymmetric (SUSY) parameter space have been extensively studied in
the literature \cite{csusy}.

The WIMPS can be detected 
on the present running or future experiments, either directly
by measuring the recoil energy when WIMP scatters off the
detector nuclei \cite{direct} or indirectly by observing the
annihilation products of the WIMPs, 
such as the antiprotons, positrons, $\gamma$-rays or 
neutrinos \cite{indirect}. 
The rate of the WIMP annihilation is proportional to the number
density square of the dark matter particles. Therefore the 
searches for the annihilation signals should aim at the 
regions with high matter densities,
such as at the galactic center \cite{gc} or the nearby subhalos
\cite{subhalo,kou,bi}. The existence of a wealth of subhalos throughout the
galaxy halos is a generic prediction of the CDM paradigm of structure
formation in the Universe. High resolution simulations show that
for CDM scenario the large scale structure forms hierarchically by continuous
merging of smaller halos and as the remnants of the merging process
about 10\% to 15\% of the total mass of the halo is in the form
of subhalos \cite{tormen98,klypin99,moore99,
ghigna00,springel01,zentner03,delucia04,kravtsov04}.
At the center of the subhalos there are high mass densities
and therefore they provide good sites for the search of WIMP 
annihilation products.

However, the analysis in Refs. \cite{flix,population} shows that 
it seems most of the
EGRET unidentified $\gamma$-ray sources are not produced by the dark
matter annihilation. In \cite{flix} by comparing the cumulative luminosity
function of subhalos and EGRET unidentified sources the authors found at
most $26\pm 11$ unidentified sources are possibly subhalos. However, location
coincidence and variability cut even exclude further of these candidates 
\cite{flix}.
In \cite{population} assuming a similar population of subhalos between
the Milky Way and M31 and
using the upper limit of $\gamma$-rays from M31 the authors found it is
highly unlikely that a large fraction of these unidentified EGRET
sources can be from subhalos. Improving the upper limit may finally
exclude this possibility \cite{population}.

The strong constraints on the possible
detectable subhalos from these works \cite{flix,population} will constrain
the properties of dark matter particles.
In the present work we assume that neutralino forms dark matter and
work in the frame of supersymmetric extension of the standard model.
Assuming that
none of the unidentified EGRET sources is from dark matter annihilation 
we will study how the properties of neutralino are constrained
resorting to the numerical simulation result of dark matter clumps 
distribution. The non-detection of subhalos finally shows a constraint
on the SUSY parameter space.

The next generation satellite based experiment, GLAST \cite{glast}, 
will greatly improve the sensitivity of EGRET. More $\gamma$-ray
sources will be detected by GLAST. Detectability of
$\gamma$-rays from the subhalos by GLAST has been studied in literature
\cite{kou,peirani,kou06,pieri,conrad,kuhlen,bertone,baltz06}.
Once such sources are detected, the follow-up study of the sources may
measure the spectrum of the annihilated $\gamma$-rays or even detect the
line emission \cite{baltz06}.  These measurement will
finally give strong implications on the properties of dark matter.
However, there are also possibilities of null result for such searches,
similar to the EGRET result or the null result of direct detection even
though the direct detection has continuously improved sensitivities.
In this case we would like to study how supersymmetry would be
constrained.

It is well known that the flux of gamma rays from the 
neutralino annihilation in a clump is given by
\begin{equation}
\label{flux}
\Phi(E)=\frac{\langle\sigma v\rangle}{2m^2}\frac{dN}{dE}
\int{dV \frac{\rho^2}{4\pi d^2}} = \frac{1}{4\pi}
\frac{\langle\sigma v\rangle}{2m^2} \frac{dN}{dE} \times \frac{1}{d^2} 
\int_{0}^{\bar{r}}4\pi r^2 \rho^2(r)dr\ ,
\end{equation}
where $\langle\sigma v\rangle$ is annihilation cross section times
relative velocity, $\frac{dN}{dE}$ is the differential flux
in a single annihilation,
$m$ is the mass of the dark matter particle,
$d$ is the distance from the detector to the source.
The flux depends on both
the distribution of the dark matter $\rho(r)$
and the particle nature of dark matter.

In the next section we first give the intensity distribution of the
subhalo $\gamma$-ray sources
according to the N-body simulation results.
Then in Sec. III we present the constraints on the SUSY parameter space
from non-detection of the subhalo $\gamma$-ray sources by EGRET and GLAST. 
We finally give a conclusion in Sec. IV.

\section{distribution of subhalos and their $\gamma$-ray intensities}

In this section we first present some simulation results about the
subhalos and then we calculate the intensities of the MW subhalos
as $\gamma$-ray sources by neutralino annihilation.

N-body simulations show that the radial
distribution of substructures is generally
shallower than the density profile of the smooth background
due to the tidal disruption of substructures
which is most effective near the galactic center \cite{diemand}.
The relative number
density of subhalos is approximately given by an isothermal
profile with a core \cite{diemand}
\begin{equation}
\label{dis}
n(r)=2n_H(1+(r/r_H)^2)^{-1}\ ,
\end{equation}
where $n_H$ is the relative number density at the scale
radius $r_H$, with $r_H$ being about $0.14$ times the halo virial radius
$r_H=0.14r_{\text{vir}}$. 
The result given above agrees well with that in another recent
simulation by Gao et al. \cite{gao}.

The differential mass
function of substructures has an approximate power law
distribution, $dn/dm\sim m^{-\alpha}$, with $\alpha = -1.7\sim -2$ \cite{diemand,gao,diemand07}.
In Ref. \cite{diemand,diemand07} both the cluster and galaxy substructure
cumulative mass functions are found to be an $m^{-1}$ power law,
$n_{\text{sub}}(m_{\text{sub}} > m)\propto m^{-1}$,
with no dependence on the mass of the parent halo.
A slight difference is found in a simulation by Gao \textit{et al.}
\cite{gao} that the cluster substructure is more abundant than galaxy
substructure since the cluster forms later and more substructures
have survived the tidal disruption.
The mass function for both scales are well fitted by $dn/dm\propto m^{-1.9}$.
Taking the power index of the differential mass function
greater than $-2$ makes the fraction of the total mass enclosed in
subhalos insensitive to the mass of the minimal subhalo we take.
The mass fraction of subhalos estimated in the literature is around
between 5 percent to 20 percent \cite{ghigna00,springel01,stoehr,diemand07}.
In this work we will take the differential index of $-1.9$ and the
mass fraction of substructures as 10 percent.


To calculate the $\gamma$-ray intensities from the dark matter clumps
we first realize a MW-like halo with a population of subhalos 
due to the subhalo distribution function from simulation, which is given
above.
The mass of the substructures are taken randomly between
$M_{\text{min}}=10^6 M_\odot$, which is the lowest substructure
mass the present simulations can resolve \cite{cut}, and the maximal mass
$M_{\text{max}}$.
The maximal mass of substructures is taken to be $0.01M_{\text{vir}}$
since the MW halo does not show recent mergers of satellites with masses
larger then $\sim 2\times 10^{10}M_\odot$.
The $\gamma$-ray flux is quite insensitive to
the minimum subhalo mass since the flux from a single subhalo
scales with its mass \cite{kou,aloisio,diemand07}.

However, due to the finite spatial resolution of the N-body simulations
the distribution in Eq. (\ref{dis}) is an extrapolation of the
subhalo distribution at large radius. The formula
underestimates of the tidal effect which destroys most substructures
near the Galactic Center (GC). We take the tidal effects into
account under the ``tidal approximation'', which assumes that
all mass beyond the tidal radius is lost in a single
orbit while keep its density profile inside the tidal radius intact.
                                                                                
The tidal radius of the substructure is defined as the radius 
at which the tidal forces of the host exceeds the self gravity of the
substructure. Assuming that both the host and the substructure gravitational
potential are given by point masses and considering the centrifugal
force experienced by the substructure the tidal radius at the Jacobi limit
is given by \cite{hayashi}
\begin{equation}
r_{\text{tid}} = r_c \left( \frac{m}{3M(r < r_c)} \right)^{\frac{1}{3}}\ ,
\end{equation}
where $r_c$ is the distance of the substructure to the GC,
$M(r < r_c)$ refers to the mass within $r_c$.

The substructures with $r_{\text{tid}} \lesssim r_s$ will be disrupted.
The mass of a substructure is recalculated by subtracting the mass
beyond the tidal radius in realizing the MW-like halos.
After taking the tidal effects into account we find 
the substructures near the GC are disrupted completely.  

\subsection{ concentration parameter }

We adopt both the NFW \cite{nfw97,nfws} and Moore \cite{moore,moores} 
profiles for the subhalos in our calculation, 
which can be written in a general form as
\begin{equation}
\rho=\frac{\rho_s}{(r/r_s)^{\gamma}[1+(r/r_s)^{\alpha}]^{(\beta
-\gamma)/\alpha}},
\label{rho}
\end{equation}
where $\rho_s$ and $r_s$ are the scale density and scale radius
respectively. $(\alpha, \beta, \gamma) = (1,3,1)$ and $(1.5,3,1.5)$ are for
the NFW and Moore profiles respectively. 

The free parameters $\rho_s$ and $r_s$ can be determined by the mass
and concentration parameter of the subhalo.
The concentration parameter is defined as
\begin{equation}
\label{con}
c=\frac{r_{vir}}{r_{-2}}\ ,
\end{equation}
where $r_{vir}$ is  the virial radius of the halo and $r_{-2}$ is the radius
at which the effective logarithmic slope of the
profile is $-2$, i.e., $\frac{d}{dr}(r^2\rho(r))\left|_{r=r_{-2}}=0\right . $.
For the NFW profile we have $r_s=r_{-2}$, while for
the Moore profile we have $r_s=r_{-2}/0.63$.

The concentration parameter is a crucial parameter in determining 
the $\gamma$ ray fluxes from subhalos.
From the definition of the concentration parameter in Eq. (\ref{con})
and the annihilation flux in Eq. (\ref{flux})
we can easily get that the annihilation flux from a clump is proportional to
$\phi \sim A \rho_s^2 r_s^3 \sim A m_{sub} c^3$ with $A$ a
flat function of $c$. Therefore the annihilation flux is very
sensitive to the concentration parameter. 
The concentration parameter is obtained by N-body simulations. 
However, due to the finite resolution of N-body simulations, 
numerical convergence has not been established, especially for the 
evolution of subhalos.
Adopting different models the predicted detectable number of subhalos 
at GLAST (at 5$\sigma$ for 1 year exposure) spans from $\lesssim 1$ by
Koushiappas et al. \cite{kou}
to at most about $300$ by Baltz et al. \cite{baltz06}. 
Recently Pieri et al. tried to
classify different cases by modeling 
the subhalos concentration
parameter and found the detectable number of subhalos at GLAST ranges
from $\sim 0$ to $\sim 40$ \cite{pieri} for different models they adopted. 
Another way for this kind of study is directly
based on the simulation result, such as in the recent work by Diemand et al.
\cite{diemand07}.
In this work we will adopt different analytic models or the fit formulas 
based on simulation results about subhalos in the literature 
to discuss their detectability. Uncertainties of the
simulation results are thus included in our study.

\begin{figure}
\includegraphics[scale=0.9]{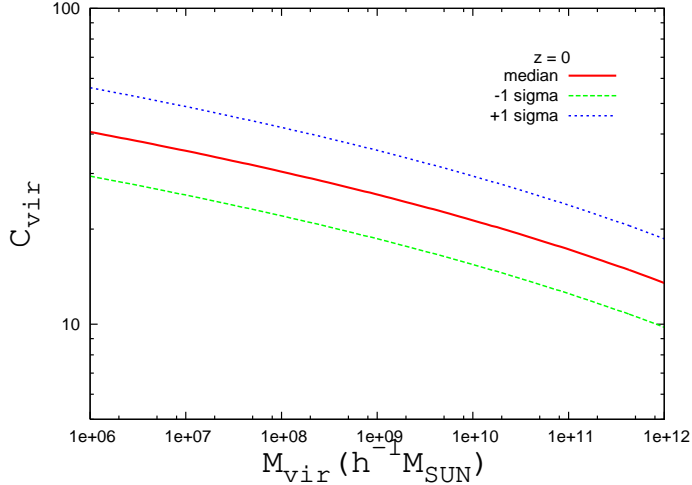}
\caption{\label{concen}
Concentration parameter as a function of the virial mass
calculated according to the Bullock model\cite{bullock01}.
The model parameters are taken as $F=0.015$ and $K=4.4$.
The cosmology parameters are taken as  $\Omega_M=0.3$, $\Omega_\Lambda=0.7$,
$\Omega_Bh^2=0.02$, $h=0.7$, $\sigma_8=0.9$ with three generations of
massless neutrinos and a standard scale invariant 
primordial spectrum.  Both the median and the $\pm 1\sigma$ values of 
the concentration parameters are plotted.
}
\end{figure}

We first introduce
a semi-analytic model to determine concentration parameter given by
Bullock et al. \cite{bullock01} 
which is built based on their simulation result.
In the model, at an epoch of redshift $z_c$ a typical 
collapsing mass $M_{*}(z_c)$
is defined by $\sigma[M_{*}(z)]=\delta_{sc}(z)$, where 
$\sigma[M_{*}(z)]$ is the linear rms density fluctuation on the
comoving scale encompassing a mass $M_*$, $\delta_{sc}$ is the
critical overdensity for collapsing  at the spherical collapse model.
The model assumes the typical collapsing mass is related to
a fixed fraction of the virial mass of a halo $M_{*}(z_c)=FM_{\text{vir}}$.
The concentration parameter of a halo with virial mass $M_{\text{vir}}$
at redshift $z$ is then determined as
$c_{\text{vir}}(M_{\text{vir}},z)=K\frac{1+z_c}{1+z}$. Both $F$ and $K$
are constants to fit the numerical simulations.
A smaller $M_{\text{vir}}$
corresponds to a smaller collapsing mass and early collapsing epoch when
the Universe is denser and therefore a larger concentration parameter.
Fig. \ref{concen}
plots the concentration parameter at $z=0$ as a function of the
virial mass of a halo according to the Bullock model\cite{bullock01}.

From Fig. \ref{concen} we can see that between the masses
$10^6 M_\odot\sim 10^{10} M_\odot$ an experiential
formula $c_{\text{vir}}\propto M_{\text{vir}}^{-\beta}$ 
reflect the simulation result accurately. We expect that this 
power law relation should
be very well followed, since subhalos form early at the epoch
when the Universe is dominated by matter with approximate
power-law power spectrum of fluctuations\cite{bullock01}.

In the literature another widely adopted semi-analytical model
for the concentration parameter is given by Eke, Navarro and Steinmetz (ENS)
\cite{eke}. We also adopt the ENS model for the
$\Lambda$CDM model with $\sigma_8=0.9$.
The other two models we adopted are the simulation results by Reed et al
\cite{reed} and that by Bullock for the subhalos in dense environment 
\cite{bullock01}.
We will show that these models predict very different annihilation fluxes.
Especially for subhalos
within the dense environment, simulation indicates it may have greater
concentration than these of isolated halos \cite{bullock01} and
therefore lead to larger annihilation flux.

\subsection{ $\gamma$-ray intensity of the subhalos  }

\begin{figure}
\includegraphics[scale=0.8]{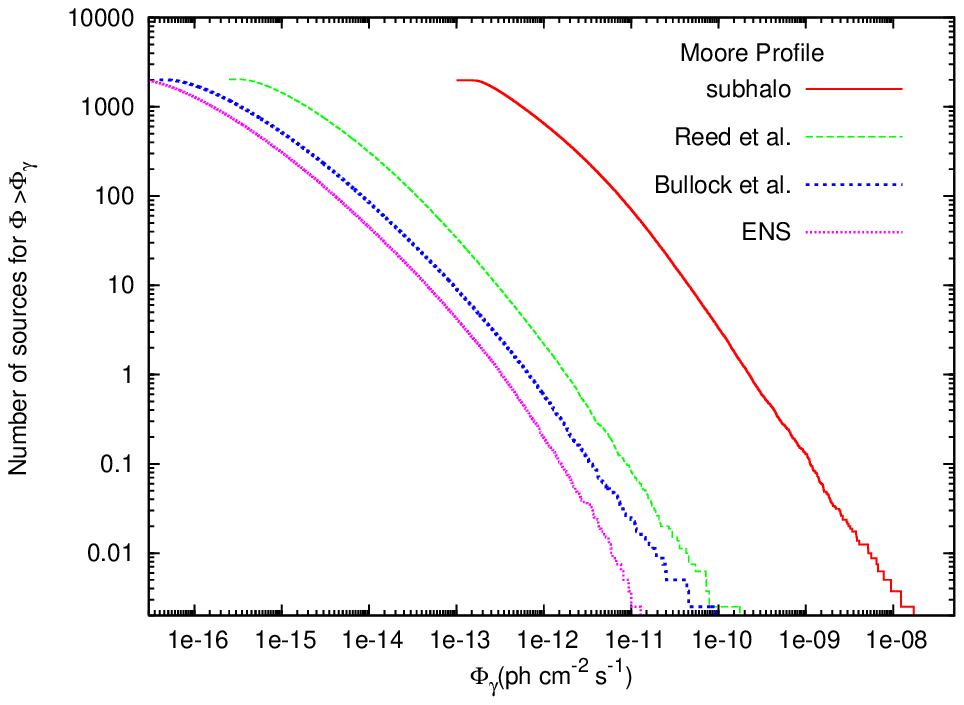}
\includegraphics[scale=0.8]{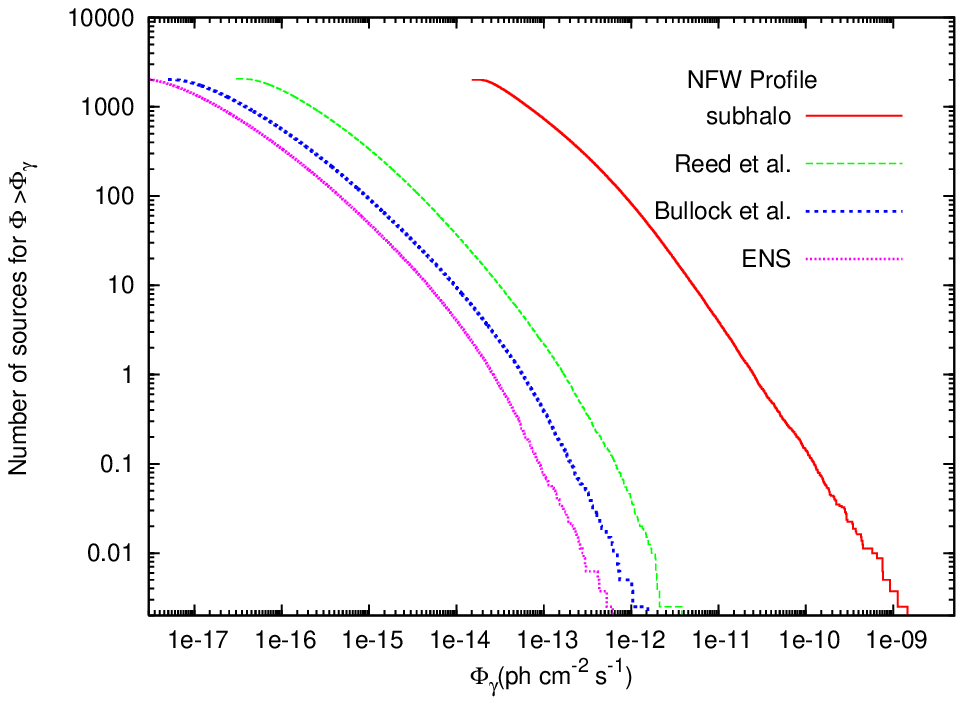}
\caption{\label{result}
The cumulative number of subhalos as function of the
integrated $\gamma$-ray fluxes $n(>\Phi_\gamma)$ for the Moore
profile (left panel) and the NFW profile (right panel)
within the solid angle of $~\pi$.
Subhalos are taken as point sources within the angular resolution
of $1^\circ$.
The curves are corresponding to different simulation results, where
`subhalo' denotes the model for subhalos within a smooth halo \cite{bullock01};
`Reed et al.' refers to the simulation results given by Reed et al in
\cite{reed};
`Bullock et al.' uses the median $c_{vir}-M_{vir}$ relation for
distinct halos of the Bullock model in \cite{bullock01};
`ENS' refers to the result of Eke et al. \cite{eke}.
}
\end{figure}

Once the profile parameters of each subhalo are determined and, 
in each realization of 
the MW-like halos, the distribution of subhalos is also known
we can calculate the $\gamma$-ray fluxes from these subhalos.
Then we can plot the number of sources as function of their intensities.
By realizing one hundred such MW sized halos
we give the averaged number of sources as a function of their intensities.
We have fixed the property of neutralino by requiring the
$\gamma$-ray flux of $3.7\times 10^{-9} \text {ph}\ cm^{-2} s^{-1} sr^{-1}$
from the GC assuming a NFW profile. Fixing flux from the GC
actually gives the relative
intensities between the GC and each subhalo.

Fig. \ref{result} gives the cumulative number of subhalos emitting
$\gamma$-rays with fluxes greater than a value $\Phi_\gamma$.
In the left
panel we plot the results for the Moore profile while the right
panel is for the NFW profile. 
From Fig. \ref{result}  we can easily read the expected number
of detectable subhalos if the sensitivity of a
detector is given with the same field of view.
For example, the sensitivity of GLAST at $5\sigma$ for $1$ year exposure
reaches $2\times 10^{-10} cm^{-2} s^{-1} sr^{-1}$ \cite{Morselli}. 
For comparison we adopt the same particle factor as Pieri et al. \cite{pieri}, which is about
$3$ orders of magnitude than our particle factor adopted here and get the
detectable subhalo number in NFW profile is from $\sim 0$ to $\sim 400$.
The maximal detectable number subhalos predicted by Pieri et al \cite{pieri}
is only about $40$, however, which is given in a different scenario from
the `subhalo' case here.
Our result is
consistent with the result by Baltz et al. \cite{baltz06}.
Even for this scenario and with the optimistic particle factor we have
only about $2$ detectable subhalos at EGRET, whose $5\sigma$ sensitivity
can only reach $\sim 10^{-8} cm^{-2} s^{-1}$.

Fig.  \ref{result} shows that there is a large discrepancy for 
predictions based on different models. Especially the `subhalo'
model gives much greater prediction. The reason is directly related
with the large concentration parameters for the subhalo scenario.
It should be noted that the other three models actually
describe distinct halos with small masses.
A qualitative simulation result about the concentration parameter is that
it is determined by the halo collapse time, as shown in the Bullock model. 
The reason of large concentration parameter for `subhalo'
is that in dense environment halos tend to collapse much earlier 
\cite{bullock01}. Tidal stripping 
may also lead to stronger mass dependence of concentration on the subhalo mass.
Another reason is that a high density environment likely leads to extreme
collapse histories of subhalos and frequent merger events which affect
the final concentration (For more discussions see \cite{bullock01}).

\subsection{detector sensitivities}

Before going on to the next section we first discuss the sensitivities
of EGRET and GLAST.
The detectability of a signal is defined by the ratio of
the signal events to the fluctuation of the background.
Since the background follows the Poisson statistics, its
fluctuation has the amplitude proportional to $\sqrt{N_B}$.
The \textit{significance} of the detection is quantified by
$\sigma=\frac{n_\gamma}{\sqrt{N_B}}$.

The signal events are given by
\begin{equation}
\label{ngamma}
n_\gamma= \epsilon_{\Delta \Omega}\int_{E_{th},\Delta \Omega} A_{eff}(E,\theta)
\phi(E) dE d\Omega dT\ \ ,
\end{equation}
where $\epsilon_{\Delta \Omega}=0.68$ is the fraction of signal
events within the angular resolution of the instrument and the
integration is for the energies above the threshold energy $E_{th}$, 
within the angular resolution of the instrument $\Delta \Omega$ and for the
observational time. Generally the effective
area $A_{eff}$ is a function of energy and zenith angle.
The $\phi(E)$ is the flux of $\gamma$-rays from DM annihilation.
We take $E_{th}= 1$ GeV for both EGRET and GLAST.
The EGRET has angular resolution of $\sim 1^\circ$ while
GLAST has much better angular resolution of $\sim 0.1^\circ$.

The corresponding
expression for the background is similar to Eq. (\ref{ngamma}).
Adopting the measured background flux $\phi^{BG}(E)$,
which is expected to get much better precision by GLAST and PAMELA,
and if we know the effective area of the detectors and the identification
efficiency for the charged particles (of hadronic and
electronic background) and photons we can get the sensitivity of the detector
\cite{bi}.
The `sensitivity' means for some time exposure, for example, for one year,
the minimal flux the source has in order to have a $5\sigma$ detection.

The sensitivity is not difficult to estimate for EGRET and GLAST, 
as given in \cite{bi}. However, a careful simulation of the detector is
beyond the present study. We will take the sensitivities of EGRET and GLAST 
directly from the literature \cite{Morselli},
that is, $3\times 10^{-8}$ and $2\times 10^{-10}$
ph cm$^{-2}$s$^{-1}$ respectively.
In \cite{Morselli} the sensitivities of EGRET and GLAST are for one
year of all sky survey with the diffuse gamma background from EGRET
as $2\times 10^{-5} ph\ cm^{-2} s^{-1} sr^{-1} (100 MeV/E)^{1.1}$,
the typical background at high galactic latitudes.
Considering background at different latitudes and longer obervation time
will certainly change the sensitivity.
However, the exact values of the detector sensitivities are not very important,
since the constraints on the SUSY parameter space given
in the following can be simply rescaled with the sensitivity.
From Fig. \ref{result} we can easily understand this: if sensitivity is
lowered by a factor $n$ the particle factor can be probed is also lowered
by the factor $n$.

For GLAST we have calculated the similar result to that in 
Fig. \ref{result} with better angular resolution. For Moore
profile there is very small difference from Fig. \ref{result}, which means most
annihilation takes place within the very small innermost region at the
halo center. For the NFW profile there are difference for the number of 
the brightest $\gamma$-ray sources, which may be from subhalos near the Sun.
For these nearby sources different angular resolution leads to different
$\gamma$-ray fluxes when taking the NFW profile.

\section{ constraints on the SUSY parameters}

From the results in last section we can predict the
number of $\gamma$-ray sources detectable in EGRET or GLAST
for any SUSY models. Conversely, if no source is detected the
SUSY models are constrained.

We will work in the frame of MSSM, the low energy
effective description of the fundamental theory at the
electroweak scale. 
By doing a random scan 
we give how the parameter space is constrained 
by these detectors. 

However, there are more than one hundred free SUSY breaking
parameters even for the R-parity conservative MSSM.
A general practice in phenomenological studies is to assume 
some simple relations between the parameters
and greatly reduce the number of free parameters.
Following the assumptions in DarkSUSY \cite{darksusy} we take
seven free parameters in calculating dark matter production and annihilation,
i.e., the higgsino mass parameter $\mu$, the wino mass parameter $M_2$,
the mass of the CP-odd Higgs boson $m_A$, the ratio of the Higgs
Vacuum expectation values $\tan\beta$, the scalar fermion mass parameter
$m_{\tilde{f}}$, the trilinear soft breaking parameter $A_t$
and $A_b$. 
All the sfermions have taken a common soft-breaking mass 
parameter $m_{\tilde{f}}$; all trilinear
parameters are zero except those of the third family; the bino and wino
have the
mass relation, $M_1=5/3\tan^2\theta_W M_2$, coming from the unification
of the gaugino mass at the grand unification scale.
The simplification of the parameters
actually does not decrease the generality of our
discussion, since the seven parameters are the most relevant ones for
our purpose. Including other parameters will not change our results much.
                                                                                
We perform a numerical random scan in the 7-dimensional
supersymmetric parameter space using the package DarkSUSY
\cite{darksusy}. The ranges of the parameters are as following:
$50 GeV < |\mu|,\ M_2,\ M_A,\ m_{\tilde{f}} < 10 TeV$,
$1.1 < \tan\beta < 61$, $-3m_{\tilde{q}} < A_t, \ A_b < 3m_{\tilde{q}}$,
$\text{sign}(\mu)=\pm 1$.
The parameter space is constrained by the theoretical consistency
requirement, such as the correct vacuum breaking pattern,
the neutralino
being the LSP and so on. The accelerator data
constrains the parameter further
from the spectrum requirement, the invisible Z-boson width and
the branching ratio of $b\to s\gamma$ \cite{darksusy}.

The SUSY models are divided into two groups: those satisfy
the constraint of dark matter relic density within 
4$\sigma$ for $\Omega_\chi h^2 = 0.105^{+0.007}_{-0.013}$ \cite{mac}
and those do not satisfy, i.e., $\Omega_\chi h^2 < 0.053$.
The effect of coannihilation between the fermions is taken into account
when calculating the relic density numerically.
For the second group of models we assume the neutralino 
is produced by some nonthermal mechanism \cite{nonthermal}
to satisfy the observation. 

\begin{figure}
\includegraphics[scale=1.2]{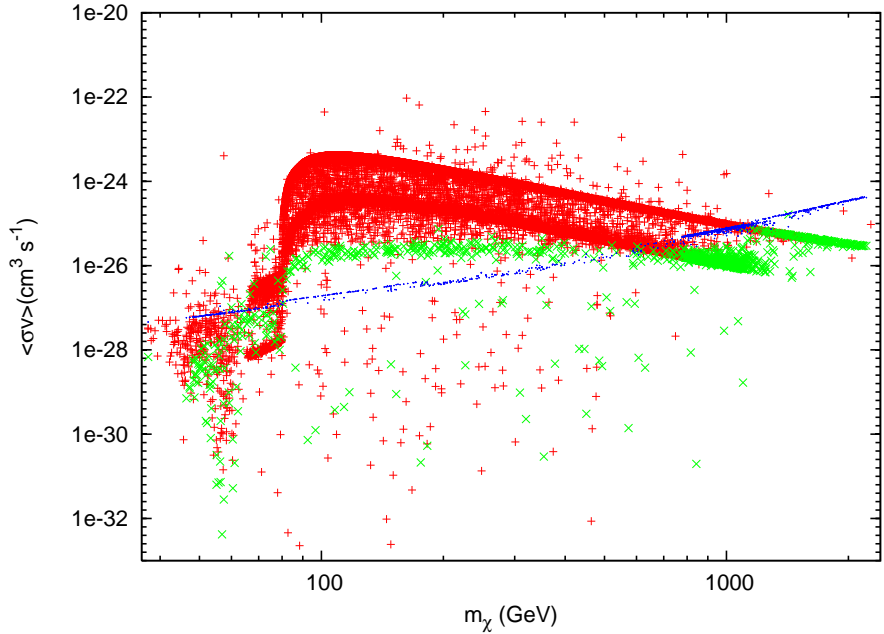}
\caption{\label{ori}
The points in the figure represent models produced randomly
in the SUSY parameter space.
The narrow strip represents the critical values of $\langle \sigma v\rangle$
for these models, as explained in the text.
In the following figures we use thick curves to represent the strip.
}
\end{figure}

\begin{figure}
\includegraphics[scale=0.8]{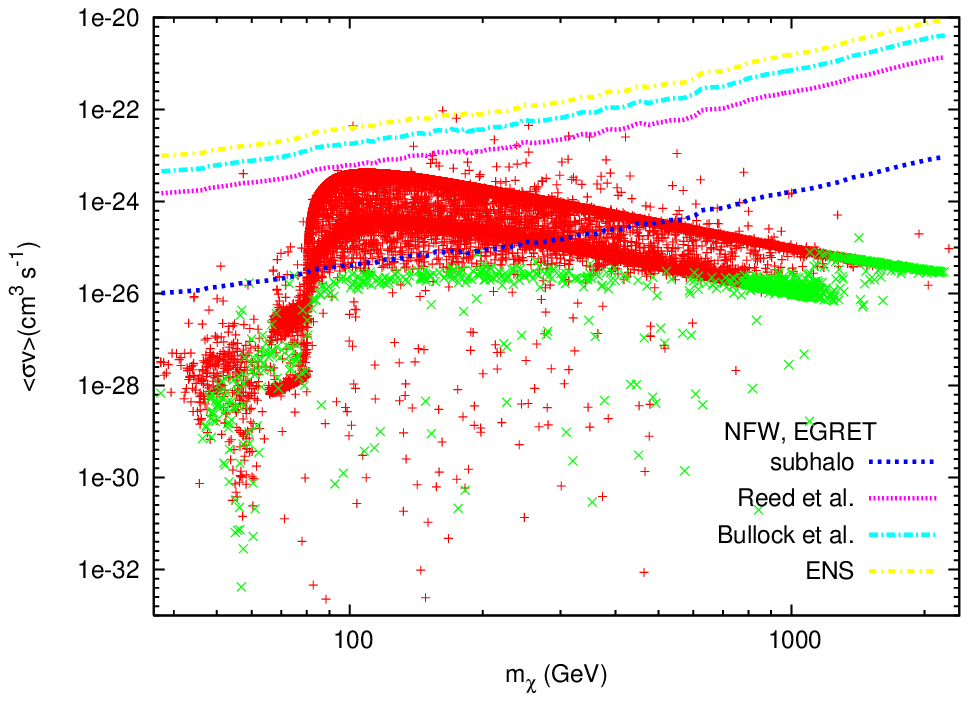}
\includegraphics[scale=0.8]{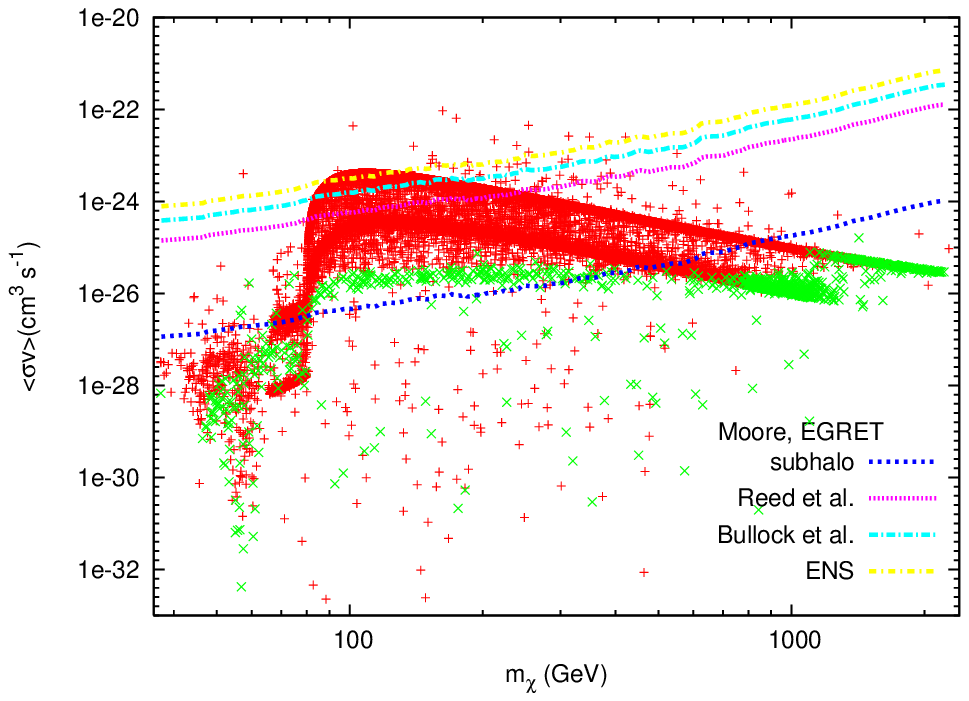}
\caption{\label{egret}
Constraints of EGRET on
the SUSY parameter space if no $\gamma$-ray sources are detected at
the 5$\sigma$ level from dark matter clumps. 
The left panel gives the constraints
assuming a NFW profile while the right is for Moore profile.
The points in the figure represent models produced randomly 
in the SUSY parameter space.
Models above the curves are ruled out.
Different curves are given adopting different simulation results, 
as explained in the text.
The models which satisfy the relic density within $4\sigma$ (green points)
have smaller
$\langle \sigma v\rangle$ than these having $\Omega h^2 < 0.053$ (red points).
}
\end{figure}

\begin{figure}
\includegraphics[scale=0.8]{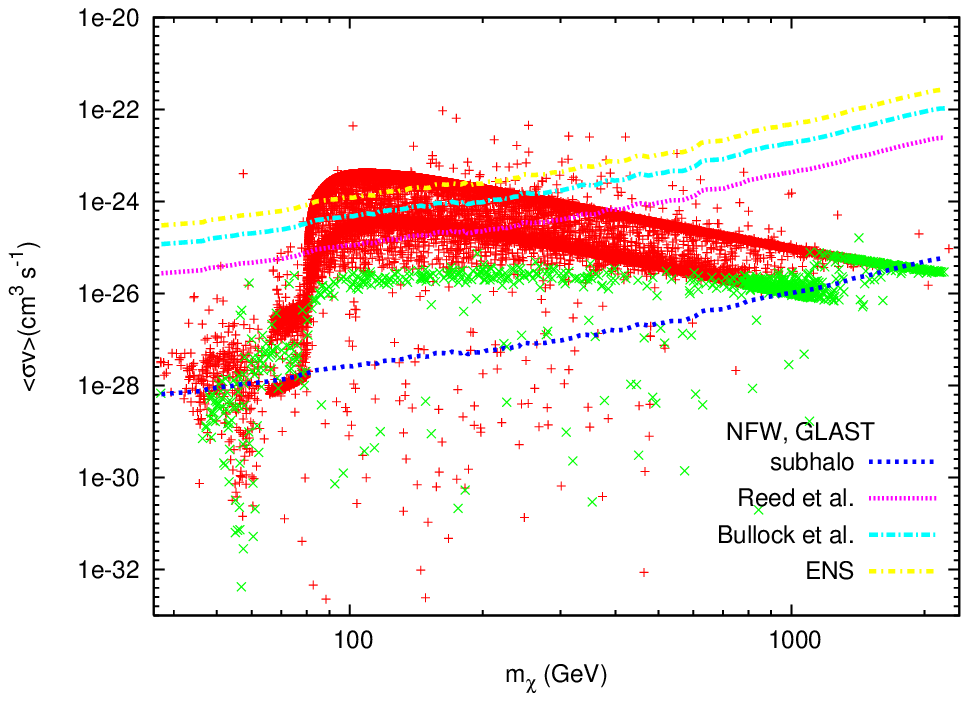}
\includegraphics[scale=0.8]{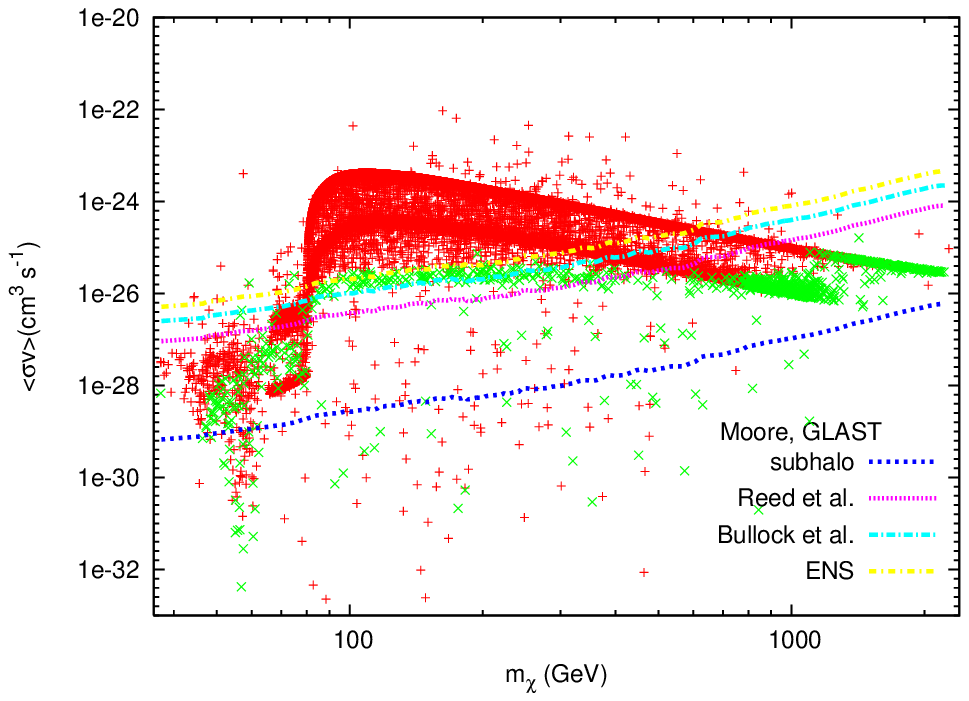}
\caption{\label{glast}
Same as Fig. \ref{egret} except that
the constraints are from GLAST.
}
\end{figure}
                                                                                
We then derive the constraints on the SUSY parameter space.
Having known the sensitivities, according to the
result in Fig. \ref{result} we can get the detectable number of $\gamma$-ray
sources in EGRET and GLAST for the SUSY model we take in last section.
When we scan in the SUSY parameter space, we calculate the average 
detectable number of subhalos at EGRET and GLAST for each SUSY model. 
Then we scale the value of 
$\langle \sigma v \rangle$ to
a critical value that only one subhalo can be detected. 
The SUSY parameters with 
larger $\langle \sigma v \rangle$ than the critical value of 
$\langle \sigma v \rangle$ should be excluded by the experiments
if null results are gotten.
In Fig. \ref{ori} we show the critical values of $\langle \sigma v \rangle$ 
according to the procedure above, which form a narrow strip. 
We would expect that the constraints should be divergent since the annihilation
final states should be very different. However, according to Fig. \ref{ori}
we know that the $\gamma$-ray spectra from different final states with same
$m_\chi$ should be quite similar so that we get convergent result. 
In the following figures we use thick curves to represent these strips. 

In Fig. \ref{egret}  we show the constraints of EGRET on
the SUSY parameter space if no unidentified $\gamma$-ray sources 
are from dark matter clumps.
The left panel gives the constraints
assuming NFW profile while
the right is for the Moore profile. 
We notice for NFW profile only the `subhalo' scenario can constrain
the non-thermal SUSY models. For Moore profile the `subhalo' scenario
can also constrain a part of the thermal SUSY models. 
The other scenarios have quite weak constraints on the models
by EGRET. 
For light neutralinos the constraints on $\langle \sigma v\rangle$ 
are $10^{-26} cm^{-3} s^{-1}$ and $10^{-27} cm^{-3} s^{-1}$ for
the NFW and Moore profiles respectively in the `subhalo' scenario.
For other models the constraints are $\sim 10^{-24} cm^{-3} s^{-1}$ 
and $\sim 10^{-25} cm^{-3} s^{-1}$ for NFW and Moore profiles respectively.
We notice that the models with  $\Omega h^2 < 0.053$ have a greater  
$\langle \sigma v\rangle$ than those thermal models and therefore
produce larger $\gamma$-ray fluxes.
These models are easier to be ruled out.

In Fig. \ref{glast} we show the similar constraints on SUSY by
GLAST, which can give a much severer constraints on the parameter space
than EGRET gives.
For the Moore profile all scenarios can put constraint on the SUSY models.
In this case a large fraction of the nonthermal parameter space will be 
ruled out. 
The constraints on $\langle \sigma v\rangle$ for light neutralinos 
are now $10^{-28} cm^{-3} s^{-1}$ and $10^{-29} cm^{-3} s^{-1}$ for
the NFW and Moore profiles respectively in the `subhalo' scenario.
For other models the constraints on $\langle \sigma v\rangle$ reach 
$\sim 10^{-26} cm^{-3} s^{-1}$
and $\sim 10^{-27} cm^{-3} s^{-1}$ for NFW and Moore profiles respectively.

\section{Discussions and Conclusion}

Since the rate of dark matter annihilation is proportional to the DM density
square, the Galactic center had been considered as the most promising site
to search for the annihilation signals. 
The possibility of detecting dark matter annihilation from the
GC has been extensively studied in literature \cite{gc}. 
However, the GC is a very complex environment. 
The dark matter density profile near the GC is complicated due
to the existence of baryonic matter and leads to difficulties in making
theoretical calculations. For example, the SMBH can
either steepen or flatten the slope of the DM profile at the innermost
center of the halo depending on the evolution of the black hole \cite{ullio}. 
Furthermore, the baryonic processes associated with
the central supermassive black hole (SMBH) and the supernova remnant
Sgr A$^*$ \cite{Zaharijas} provide a strong $\gamma$-ray background,
which has been detected by HESS \cite{gcg}, to the signals of dark matter 
annihilation
and make the detection very difficult \cite{Zaharijas}.
In \cite{Zaharijas} it is found that in case of NFW profile
GLAST can probe $\langle \sigma v\rangle$ between $ 10^{-26} cm^{-3} s^{-1}$
and $ 10^{-28} cm^{-3} s^{-1}$ for light neutralinos, which
is similar to the sensitivity by observing subhalos as shown
in Fig. \ref{glast}. However, considering
the HESS detected $\gamma$-ray background only models with 
$\langle \sigma v\rangle \gtrsim 10^{-27} cm^{-3} s^{-1}$ can be probed 
\cite{Zaharijas} from the GC by GLAST. Therefore it becomes less sensitive 
to probe the GC than detecting subhalos now.
For the case of Moore profile, sensitivity from GC observation
is improved by
two orders of magnitude, while only one order of magnitude
improvement from subhalo observation. This means similar sensitivities
will be achieved by observing the GC and subhalos.

On the contrary, subhalos provide a clean environment to search for the
annihilation signals.
Especially recent simulation shows that the DM profiles may not be universal.
Smaller subhalos may have steeper central cusp \cite{reed,jing}.
Reed et al. gave the cusp index $\gamma = 1.4 - 0.08\log(M/M_*)$ 
for halos of $0.01 M_*$ to $1000 M_*$ with a large scatter.
In this case, if taking the GC the NFW profile and the
subhalos the Moore profile, the $\gamma$-ray fluxes from the
subhalos may even be greater than that from the GC.

We expect
these sources can be detected by the satellite based experiments,
such as EGRET and GLAST. Once such sources are detected we can learn
a lot about the nature of dark matter particles by studying its luminosity
and spectrum of the annihilation.
However, study shows that most of the EGRET unidentified
$\gamma$-ray sources should not be of dark matter origin 
\cite{flix,population}.
In this work we study how the EGRET and GLAST can constrain the SUSY
models if none of the subhalo $\gamma$-ray sources are detected in
the two experiments.

We first realize one hundred MW-like halos with 
subhalos whose distribution is given according to 
the N-body simulation results. 
In each realization we calculate the $\gamma$-ray flux from the
subhalos by fixing the particle factor. 
Then we give the average cumulative number of the subhalo 
$\gamma$-ray sources as function of their flux intensities.
Once the sensitivity of detectors, such  EGRET and GLAST, are known,
we know the detectable number of this kind of $\gamma$-ray sources.
By requiring the detectable number smaller than $1$ at EGRET and GLAST
we put a constraint on the SUSY parameter space. 
Our result shows that the EGRET has already given
a moderate constraint on the SUSY parameter space if we assume none
of the unidentified  $\gamma$-ray sources are from subhalos. 
The GLAST can greatly enhance the
constraints. However, a large uncertainty comes from the simulation,
especially the property of subhalos in a dense environment.
Convergence of the subhalos property in the future high resolution
simulation will lead to more precise constraint on the
nature of dark matter particles.

\begin{acknowledgments}
This work is supported by the NSF of China under the grant
Nos. 10575111, 10773011 and supported in part by the Chinese Academy of
Sciences under the grant No. KJCX3-SYW-N2.
\end{acknowledgments}

\end{document}